\documentclass[12pt]{article}
\usepackage{amsmath}
\usepackage{amsfonts}

\numberwithin{equation}{section}
\newcommand{\R}{\mathbb{R}}
\newcommand{\C}{\mathbb{C}}
\newcommand{\E}{\mathbb{E}}
\newcommand{\p}{\partial}
\newcommand{\del}{\delta}
\begin{document}
\begin{titlepage}
\begin{center}
{\Large \bf Noncommutative Generalized
NS \\ and Super Matrix KdV Systems\\from a\\
Noncommutative Version of\\(Anti-)Self-Dual Yang-Mills Equations}
\vskip 5em
{\large M. Legar\'e}\\
\vskip 1em
Department of Mathematical Sciences\\
University of Alberta\\
Edmonton, Alberta, Canada, T6G 2G1
\end{center}
\vskip 5em

\begin{abstract}
A noncommutative version of the (anti-) self-dual Yang-Mills 
equations is shown to be related via dimensional reductions to
noncommutative formulations of the generalized ($SO(3)/SO(2)$)
nonlinear Schr\"odinger (NS) equations, of the super- Korteweg -
de Vries (super-KdV) as well as of the matrix KdV equations.
Noncommutative extensions of their linear systems and bicomplexes
associated to conserved quantities are discussed.
 
\noindent\sloppy{{\bf PACS} : 11.10.Lm, 11.15. q, 11.30. j,
11.30.Pb}

\noindent\sloppy{{\bf Keywords} : Supersymmetry, noncommutative,
integrable, self-dual Yang-Mills}
\end{abstract}
\end{titlepage}

\pagebreak

\section{Introduction}

Noncommutative geometry has recently been involved in a 
noncommutative version of gauge theory \cite{SW} related to
strings and has been stimulated by different works on field
theories defined over noncommutative spaces (for example
\cite{MSSW,GMS,P1} and references therein). Naturally, classical
integrable models have also been generalized to noncommutative
spaces (\cite{DM1,DM2,DM3} and references therein), and, for
instance, noncommutative versions of the Toda, nonlinear
Schr\"odinger (NS) and Korteweg - de Vries (KdV) equations have
been formulated. Some properties of these ``deformed" versions
have also been shown. Using bicomplexes, an infinite set
of conserved quantities has been found, which as well suggest
the complete integrability of these modified systems. In view of
the results on the deformation of the ADHM construction \cite{NS}
and its twistor interpretation \cite{KKO}, a formulation of
(anti-)self-dual Yang-Mills (abbreviated below (A-) SDYM)
equations on noncommutative spaces has been presented
\cite{T1}. Many twistor and integrability properties were shown to
be preserved in this setting. The ``deformation" equations are
simply obtained by substituting the product of fields with a
Moyal product in the classical form of the (A-) SDYM equations.
Dimensional reductions to the (noncommutative) principal chiral
field model and Hitchin equations are discussed in \cite{T1}, and
integrability properties inherited from the (A-) SDYM equations.

In this short article, dimensional reductions of a noncommutative 
version of the (A-) SDYM equations on noncommutative Euclidean
($\E^4$) and pseudo-Euclidean ($\E^{(2,2)}$) spaces are studied
with also the help of conditions on the gauge fields. First,
generalized nonlinear Schr\"odinger (NS) equations and matrix KdV
equations on noncommutative spaces are derived with associated
linear systems (or Lax pairs). The integrability of these systems
is suggested from the (anti-) self-dual Yang-Mills equations,
the presence of an infinite set of conserved quantities, and
bicomplexes, which are themselves linked to the reduced linear
systems. A noncommutative version of a supersymmetric KdV system
is also derived in a similar manner. Some of these results are an
adaptation of certain methods used in ref. \cite{LL} and
references therein for closely related problems, i.e. for the
commutative version of the systems presented below. It relates to
the approaches of refs \cite{DM2,DM3} and \cite {T1}. However, it
does not appear that general commutative reductions could be
extended to noncommutative versions in the same manner.

Section 2 introduces the notation and different elements needed
for the description of systems on noncommutative space, such as
Lax pairs or linear systems on noncommutative Euclidean and
signature zero pseudo-Euclidean spaces. Then in section 3, the
linear systems, bicomplexes, and generalized nonlinear
Schr\"odinger equations are obtained as dimensional reductions
accompanied with specific gauge fields forms (or ansatzes) of the
corresponding (A-) SDYM equations and their linear system on an
identical noncommutative space. Section 4 deals with the matrix
KdV equations, and also presents a noncommutative version of the
supersymmetric (matrix) KdV equations, again as reductions of the
SDYM equations and their linear system(s) on the appropriate
noncommutative space. Finally, section 5 suggests developments
and comments on the previous sections.

\section{Bicomplexes, Linear Systems,\\and Noncommutative
Formulations}

Definitions and applications of bicomplexes can be found for
example in the following refs \cite{L1,SML,AF,DM4}. For our
purposes, let us use the following definition below
(see for example \cite{DM4}). A bicomplex corresponds to a linear
space over $\R$ or $\C$, here denoted $V$, endowed with a grading
over the non-negative integers, i.e. $$V =
\bigoplus_{i \geq 0} V^i,$$ and two (linear) maps (operators) $d$
and $\del$ between successive spaces $V^i$ and $V^{i+1}$ in other
words, $ d : V^i \rightarrow V^{i+1}$, and $\del : V^i \rightarrow
V^{i+1},$ such that :
\begin{equation}
d^2 \cdot = 0,\quad \del^2 \cdot = 0, \quad (\del d +d\del) \cdot
= 0,
\end{equation}
where $\cdot$ stands for an element of $V$.

A set of bicomplexes can be related to linear systems of the
(A-)SDYM equations. For instance, the (A-)SDYM equations on
4-dimensional Euclidean space ($\E^4$) have the following linear
systems \cite{W1}:
\begin{align}\label{ls-1}
[D_1 + iD_2 - \lambda (D_3 \pm i D_4)] \Psi(x,
\lambda,\bar\lambda) &= 0 \nonumber\\
[D_3 \mp i D_4 +\lambda (D_1 - iD_2)] \Psi(x,\lambda,\bar\lambda)
&= 0,\\
\p_{\bar\lambda} \Psi(x,\lambda,\bar\lambda) &= 0 \nonumber,
\end{align}
where the lower sign in the above equations correspond to the
A-SDYM formulation, $\lambda \in \C P^1$, $D_\mu = \p_\mu +
A_\mu$, and $\Psi$ is a $\C$-valued column vector. On the
4-dimensional pseudo-Euclidean space of signature $0$
($\E^{(2,2)}$) with diagonal metric $(+,+,-,-)$, one finds the
following set of linear equations \cite{IP1,IP2} :
\begin{align}\label{ls-2}
[D_1 + iD_2 + \lambda (D_3 \mp i D_4)] \Psi(x,
\lambda,\bar\lambda) &= 0 \nonumber\\
[D_3 \pm i D_4 +\lambda (D_1 - iD_2)] \Psi(x,\lambda,\bar\lambda)
&= 0,\\
\p_{\bar\lambda} \Psi(x,\lambda,\bar\lambda) &= 0 \nonumber,
\end{align}
where here too, the lower sign applies to the A-SDYM equations,
and $\lambda \in $ (a sheet of hyperboloid ${\mathbb{H}}^2$).

The compatibility equations of the linear systems (\ref{ls-1})
and (\ref{ls-2}) are, respectively, the (A-)SDYM equations on
$\E^4$ or
$\E^{(2,2)}$.

Note that, for simplicity in later calculations, the A-SDYM
equations on $\E^4$ could be transformed to \cite{T1,AC} :
\begin{equation}
F_{z_1z_2} = 0, \quad F_{\bar z_1\bar z_2} = 0, \quad F_{z_1\bar
z_1} + F_{z_2\bar z_2}  = 0,
\end{equation}
where : $F_{z_iz_j} = \p_{z_i} A_{z_j} - \p_{z_j} A_{z_i} +
[A_{z_i}, A_{z_j}]$, and $F_{z_i\bar z_j} = \p_{z_i} A_{\bar z_j}
- \p_{\bar z_j} A_{z_i} + [A_{z_i}, A_{\bar z_j}]$, using the
following change to null variables : $z_1 = x_3+ix_4, z_2 =
x_1+ix_2, \bar z_1 = x_3 - ix_4, \bar z_2 = x_1 -i x_2,$ with
$i,j=1,2.$

Accordingly, the associated linear system becomes :
\begin{align}\label{ls-3}
[(\p_{\bar z_1} - \lambda \p_{z_2}) + (A_{\bar z_1} - \lambda
A_{z_2})] \Psi (z_i,\bar z_j, \lambda, \bar\lambda) &=
0,\nonumber\\
[(\p_{\bar z_2} + \lambda \p_{z_1}) + (A_{\bar z_2} + \lambda
A_{z_1})] \Psi (z_i,\bar z_j, \lambda, \bar\lambda) &=
0,\\
\p_{\bar\lambda} \Psi(z_i, \bar z_j,\lambda,\bar\lambda) &= 0
\nonumber.
\end{align}

As for the SDYM equations on $\E^{(2,2)}$, a change to null
variables \cite{LL,IP1,IP2} :
\begin{align}
t&=\dfrac{1}{\sqrt{2}} (x^2-x^4)&,& &y=\dfrac{1}{\sqrt{2}}
(x^1-x^3),\nonumber\\
u&=\dfrac{1}{\sqrt{2}} (x^2+x^4)&,& &z=\dfrac{1}{\sqrt{2}}
(x^1+x^3),
\end{align}
leads to the following corresponding linear system on
$\E^{(2,2)}$ :
\begin{align}\label{ls-4}
(D_z + \omega D_u) \Psi(x,\omega,\bar\omega) = 0,\nonumber\\
(D_t - \omega D_y) \Psi(x,\omega,\bar\omega) = 0,\\
\p_{\bar\omega} \Psi(x,\omega,\bar\omega) = 0, \nonumber
\end{align}
where the parameter : $\omega =
i\dfrac{(1-\lambda)}{(1+\lambda)}$, and : $D_t =
\dfrac{1}{\sqrt{2}}(D_2-D_4), D_u = \dfrac{1}{\sqrt{2}}(D_2+D_4),
D_y = \dfrac{1}{\sqrt{2}}(D_1-D_3), D_z = \dfrac{1}{\sqrt{2}}
(D_1+D_3)$.

However, if one builds a bicomplex based on the previous type of
linear systems (\ref{ls-1}, \ref{ls-2},\ref{ls-3}) with parameter
$\lambda$ :
\begin{align}\label{ls-5}
\mathcal{D}_1 \Psi &= [\mathcal{O}_1 + \lambda
\mathcal{O}_1^\lambda ] =0, \nonumber\\
\mathcal{D}_2 \Psi &= [\mathcal{O}_2 + \lambda
\mathcal{O}_2^\lambda ] =0,
\end{align}
as the set of two operators $d$ and $\del$ on $\Psi : \R^4
\rightarrow \C^n, \in V^0$ \cite{DM1,DM2} :
\begin{align}\label{bicomplex}
d\Psi &= \mathcal{O}_1 \Psi \xi_1 + \mathcal{O}_2 \Psi
\xi_2,\nonumber\\
\del\Psi &= \mathcal{O}_1^\lambda \Psi \xi_1 +
\mathcal{O}_2^\lambda \Psi \xi_2,
\end{align}
or, in short, $d\Psi + \lambda \del \Psi = 0$ , which
resembles the ``linear equation " formulation of
\cite{DM2,DM3}, where $\xi_1,\xi_2 \in \Lambda^1$, (which can be
simply extended to $V$,) then the conditions for these
operators to form a bicomplex : $d^2 = 0, \del^2 = 0, d\del +
\del d = 0$, correspond exactly to the compatibility or
integrability conditions of the linear system (\ref{ls-5}), and
provide the (A-)SDYM equations on the respective space for the
linear systems (\ref{ls-1}), (\ref{ls-2}), (\ref{ls-3}). One notes
that the system (\ref{bicomplex}) is as well invariant under gauge
transformations. 

Now, it would of interest to introduce a noncommutative version of
the four dimensional pseudo-Euclidean space of signature $0$
($\E^{(2,2)}$). Let us assume for the coordinates
$x^\mu, \mu,\nu = 1,2,3,4$, the following commutation relations :
\begin{equation}
[x^\mu,x^\nu] = i\theta^{\mu\nu},
\end{equation}
where the quantities $\theta^{\mu\nu}$ are real constants. The
associative product on the space of functions over $\E^{(2,2)}$
is then substituted here by the (associative, but
noncommutative) Moyal product \cite{M1}, denoted $\ast$, of two
functions $f$ and $g$ on $\E^{(2,2)}$ :
\begin{equation}
(f\ast g)(x) = \exp[\sum_{\mu,\nu =1}^4
\dfrac{i}{2}\theta^{\mu\nu} \p_{x^\mu}\p_{\tilde x^\nu}]
f(x^\lambda)g(\tilde x^\sigma) \arrowvert_{x^\mu=\tilde x^\mu}.
\end{equation}

A simple noncommutative version of the (A-)SDYM equations on
$\E^4$ or $\E^{(2,2)}$ can then be obtained by using the above
$\ast$-product instead of the usual commutative product of two
functions. Thus :
\begin{equation}
F_{\mu\nu} = \p_\mu A_\nu - \p_\nu A_\mu + A_\mu\ast A_\nu -
A_\nu\ast A_\mu = \p_\mu A_\nu - \p_\nu A_\mu + [A_\mu,\kern
-6pt ^\ast A_\nu], 
\end{equation}
stands for the field strength of the gauge field components
$A_\mu$. The (A-) SDYM equations are then invariant under the
gauge transformations :
\begin{equation}
A_\mu' = g^{-1} \ast A_\mu \ast g + g^{-1}\ast \p_\mu g,
\end{equation}
with $g^{-1}$ the inverse of $g$, such that $g^{-1}\ast g = 1$.
Note that $\p_\mu$ is still a ``derivation" on the noncommutative
spaces $\E^4$ or $\E^{(2,2)}$, and that the gauge group has
not been specified yet, but such gauge theories have been
explored for $U(n)$ as gauge group \cite{T1}. Also, it is noticed
that this version differs from versions of noncommutative
theories found for instance in refs \cite{MSSW} and
\cite{GMS}.

In the following sections, slight modifications of known
reductions of the (commutative) (A-)SDYM equations to (classical)
integrable systems will be used to derive (noncommutative)
versions of the same integrable models, via translational
reductions of the (A-)SDYM equations. Corresponding reduced
noncommutative linear systems will also be obtained, and their
integrability associated to the compatibility of the original
(A-)SDYM equations.

\section{Generalized NS Equations}  

As presented in refs \cite{AC} and \cite{MS}, the NS equations
can be derived from the SDYM equations on $\E^{(2,2)}$ using
translational invariance along $z_2$ and $z_1-\bar z_1$, with the
``ansatz" :
\begin{xalignat}{2}
A_{z_1} &= 0\quad & A_{\bar z_1} &= \left [
\matrix 0 & -q \\ r & 0 \endmatrix \right ]\nonumber\\
A_{z_2} &= -\kappa \left [\matrix 1 & 0 \\ 0 & -1 \endmatrix
\right ]
\quad & A_{\bar z_2} &= -\dfrac{1}{2\kappa} \left [ \matrix q\ast
r & q_x \\ r_x & -r\ast q \endmatrix \right ].
\end{xalignat}
The residual linear system (\ref{ls-3}) takes the form: 
\begin{align}
[\p_x + (A_{\bar z_1} - \lambda A_{z_2})] \ast \Psi &=
0,\nonumber\\
[\p_t +\lambda \p_x + A_{\bar z_2}] \ast \Psi &= 0,\\
\p_{\bar\lambda} \Psi &= 0,\nonumber 
\end{align}
and then have the noncommutative generalized NS equations
given below as compatibility equations :
\begin{align}\label{nc-kdv}
2\kappa\, q_t &= q_{xx} + 2\,q\ast r\ast q,\nonumber\\
2\kappa\, r_t &= -(r_{xx} + 2\,r\ast q\ast r),
\end{align}
where : $x=z_1 + \bar z_1$, $t=\bar z_2$, and $\kappa$ is a
constant.

The equations (\ref{nc-kdv}) coincide with the equations
obtained in ref. \cite{DM2} from an almost similar bicomplex, with
$q=\bar r$ and $\kappa = \dfrac{i}{2}$. Let us add that conserved
quantities for this noncommutative system would be derivable in a
manner similar to the approach found in ref. \cite{DM2}.

\section{Super Matrix KdV Equations}

In a follow up to the work of refs (\cite{MS}) and (\cite{BD})
on the reduction of the (A-)SDYM equations to the
(commutative) KdV equation, the (commutative) matrix KdV
equations were deduced from the same original (A-)SDYM
equations via translational symmetries \cite{IP1,IP2}, and
then using Lie superalgebra valued gauge fields, a
supersymmetric version of the matrix KdV model was exhibited
\cite{LL}.

The symmetry reductions applied in refs \cite{AC} and
\cite{MS} have not allowed us to derive a noncommutative
form of these equations via the same procedure used on
noncommutative (A-)SDYM equations. Instead, the formulations
of refs \cite{LL}, \cite{IP1}, \cite{IP2}, and \cite{BD}
have been found more suitable for this purpose. Indeed,
starting from the linear system (\ref{ls-4}) and imposing
translational symmetries along the coordinates $u$ and
$y-z$, one finds the residual linear equations :
\begin{align}\label{ls-6}
[\p_t + A_t +\omega (A_z-A_y) +\omega^2 A_u]\ast \Psi &=
0,\nonumber\\
[\p_x + A_z +\omega A_u ] \ast\Psi &= 0,
\end{align}
Then, the additional expressions of the gauge components with
values in a real form of $sl(n,\C) \otimes \mathcal{A}$,
where $\mathcal{A}$ identifies the set of functions on
noncommutative $\E^{(2,2)}$, are introduced :
\begin{xalignat}{2}
A_u(t,x) &=\left [ \matrix 0_n & 0_n \\ -1_n & 0_n \endmatrix
\right ] & \quad A_z(t,x) &= \left [ \matrix 0_n & 0_n \\ U_n(t,x)
& 0_n ,\endmatrix \right ]\nonumber
\end{xalignat}
\begin{align}\label{gauge-ansatz}
A_{z-y}(t,x) &\equiv A_z(t,x)-A_y(t,x) = \left [ \matrix 0_n &
0_n \\ 0_n & 0_n \endmatrix \right ],\\
A_t(t,x) &= \left [ \matrix 0_n & 0_n \\ 3 U_n \ast U_n +
U_{n,xx} & 0_n \endmatrix \right ], \nonumber  
\end{align}where the subindex $n$ indicates the dimension
of the matrix involved, i.e. $n \times n$.

One can mention that the formulation of ref. \cite{BD} can
also provide the same noncommutative version of the KdV equations,
which arise as the compatibility of the above linear system
(\ref{ls-6}) with components (\ref{gauge-ansatz})  :
\begin{equation}
U_{n,t} = 3(U_{n,x}\ast U_n + U_n\ast U_{n,x}) + U_{n,xxx},
\end{equation}
which provides when $n=1$ :
\begin{equation}
U_t = 3(U_x\ast U + U\ast U_x) + U_{xxx},
\end{equation}
originally presented in ref \cite{DM3} as a noncommutative
version of the KdV equation, but with a different path. An
infinite set of conserved densities can be derived using a
noncommutative version of the transformation presented in
\cite{MGK} : $U = W +\lambda W_x + \lambda^2 W\ast W$ \cite{DM3}. 

On the other hand, a noncommutative version of a supersymmetric
KdV equation can also be produced from the linear systems
(\ref{ls-6}) by inserting the following ansatz for the
gauge field components into the compatibility equations :
\begin{xalignat}{2}
A_u &= \left [ \matrix 0 &0&0 \\ -1&0&0 \\ 0&0&0 \endmatrix 
\right ] \quad & A_z &= \left [ \matrix 0&0&0 \\ U&0&0
\\0&0& \theta \phi \endmatrix \right ]\nonumber \\
A_{z-y} &= A_z-A_y = \left [ \matrix 0&0&0 \\ 0&0&\theta \\
\theta&0&0 \endmatrix \right ] \quad & A_y &= \left [ \matrix
0&0&0 \\ U&0&-\theta \\ -\theta&0&\theta\phi \endmatrix \right ]
\end{xalignat}
\begin{equation}
A_t = \left [ \matrix 0&0&0 \\ 3 U\ast U + U_{xx} -\dfrac{3}{2}
\phi\ast\phi_x + \dfrac{3}{2} \phi_x\ast \phi &0&0\\
0&0& \theta (\phi_{xx} + \dfrac{3}{2} \phi\ast U
+\dfrac{3}{2} U\ast \phi) \endmatrix \right ]\nonumber
\end{equation}

The reduced equations SDYM equations using the above fields with
values in the Lie superalgebra $gl(2n/n)$, where $\theta$ is an
odd Grassmann variable, and with $U$ and $\phi$ being respectively
even and odd degree variables depending on $x$ and $t$, have
the form :
\begin{align}\label{nc-superkdv}
U_t &= 3 U_x\ast U + 3 U\ast U_x + U_{xxx} - \dfrac{3}{2}
\phi\ast\phi_{xx} + \dfrac{3}{2} \phi_{xx}\ast\phi, \nonumber \\
\phi_t &= \phi_{xxx} + \dfrac{3}{2} \phi_x\ast U + \dfrac{3}{2}
\phi \ast U_x +\dfrac{3}{2} U_x\ast \phi +\dfrac{3}{2} U\ast
\phi_x
\end{align}

It can be verified that these noncommutative equations are left
invarinat under the following supersymmetry transformations,
induced by the odd Grassmann parameter $\epsilon$ :
\begin{equation}
\del_\epsilon U = \epsilon \phi_x ,\quad\text{and}\quad
\del_\epsilon \phi = \epsilon U.
\end{equation}

A derivation of a noncommuatative formulation of supersymmetric
matrix KdV equations is similar and the resulting equations can
be written by adding a subscript $n$ to the variables $U$ and
$\phi$ in the equation (\ref{nc-superkdv}) above. 

\section{Summary/Conclusion}    

This paper has shown a relation via the procedure of reduction
\cite{O1,BK,W2} for translations between a noncommutative version
of (anti-) self-dual Yang-Mills equations and noncommutative
formulations of diverse integrable systems, the generalized NS
and matrix KdV equations, as well as a supersymmetric  integrable
model, the super matrix KdV system. It could be seen as an
extension of results published by ref. \cite{MS} and ref.
\cite{LL} in the direction of noncommutative theories. For each
of these noncommutative versions of integrable models, a
corresponding noncommutative linear system has been exhibited,
and a link to bicomplexes provided. Conserved densities would be
obtainable in a similar fashion to the cases presented by refs
\cite{DM2,DM3}.

Many directions can then be followed as future developments. One
may want to explore the set of solutions of these noncommutative
models, and more explicitly, look at the possibilities for
solitons, or related solutions in the $0$ limit of $\theta$ (see
\cite{DM3}). The integrability, Hamiltonian, and reduced twistor
interpretations could also be probed, as well as further
reductions to other integrable equations such as (2+1) systems
\cite{DM5}, using varied constraints and symmetries. Moreover,
other formulations of noncommutative gauge theories could be
examined in similar manners through a reduction process.

\section{Acknowledgements}

The author acknowledges for the support of this work a grant from 
the National Sciences Research Council (NSERC) of Canada.

\end{document}